# Colossal resistivity change besides magnetoresistance: an extended theoretical framework for electronic transport of manganites

S. Dong[1], K. F. Wang[1], Han Zhu[2], X. Y. Yao[1], and J. –M. Liu[*,1,3]

[1]*Laboratory of Solid State Microstructures, Nanjing University, Nanjing 210093, China*
[2]*Department of Physics, Princeton University, Princeton, New Jersey 08544, USA*
[3]*International Center for Materials physics, Chinese Academy of Sciences, Shenyang, China*

[Abstract] Current theoretical approaches to manganites mainly stem from magnetic framework, in which the electronic transport is thought to be spin-dependent and the double exchange mechanism plays a core role. However, quite a number of experimental observations can yet not be reasonably explained. For example, multiplicate insulator-metal transitions and resistivity reduction induced by perturbations other than magnetic field, such as electric current, are not well understood. A comprehensive analysis on earlier extensive studies is performed and two types of origins for resistivity change are highlighted. Besides the insulated-to-metallic transition induced by external field such as magnetic field, the insulated-to-insulated transition induced extrinsically is even a more important source for the colossal resistivity change. We propose an extended framework for the electronic transport of manganites, in which the contribution of charge degree of freedom is given a special priority.

PACS numbers: 75.47.Lx, 75.47.Gk, 71.30.+h

---

[*] Corresponding author. E-mail: liujm@nju.edu.cn

Manganites, typical strongly correlated electron systems, have been extensively studied over the last decade because of their colossal magnetoresistance (CMR) effect.[1] The general chemical composition for manganites is $T_{1-x}D_xMnO_3$, with T a trivalent rare earth cation, and D a divalent alkaline earth cation or $Pb^{2+}$. In mixed-valent manganites, the hopping of 3d $e_g$ electrons causes the ferromagnetic coupling between localized Mn 3d $t_{2g}$ spins (so called double exchange or DE mechanism). Quite a number of theoretical approaches were proposed to explain the CMR effect and relevant phenomena, in which the DE mechanism is thought to be in a core position, with additional interactions such as Hund coupling, Jahn-Teller effect, Coulomb repulsion and antiferromagnetic super-exchange included.

A number of previous investigations unveiled that manganites are intrinsically inhomogeneous. Phase separation (PS) and percolation were observed by experiments [2-4] and subsequently confirmed theoretically.[5-9] The current theories originate from three fundamental points: first, because of DE mechanism, ferromagnetic (FM) phase is metallic while antiferromagnetic (AFM) phase and paramagnetic (PM) phase are insulated. Second, the PS consists of FM metallic clusters and other insulated regions. It is argued that the PS on micrometer scale is induced by quenched disorders, in which the percolation of metallic cluster embedded in insulated matrix may occur. Finally, some insulated regions can be converted into metal upon application of external magnetic field, which is characterized by the raise of temperature $T_{IM}$ for insulator-metal transition (IMT) and enhanced magnetoresistance near $T_{IM}$. In accordance, the Curie temperature ($T_C$) is raised too.

Indeed, the magneto-transport behaviours of those large-bandwidth manganites, e.g. $La_{1-x}Sr_xMnO_3$ ($x$~0.3),[2] can be reasonably described by the DE framework. Some other manganites, however, offer anomalous properties from the DE framework. First, most manganites present charge ordered (CO) states in a broad doping density region over a finite temperature ($T$) range. Although CO states may be melted by magnetic field,[10] a large magnetic field threshold is needed, beyond which an appreciable change of resistivity becomes possible. Such threshold for $Pr_{0.6}Ca_{0.4}MnO_3$ is about 4.2T at 30K.[11] Quite different behaviours were observed for $La_{2/3}Sr_{1/3}MnO_3$ single crystals, whose resistance response to magnetic field is almost linear.[12] Second, for some manganites of not very large bandwidth, e.g. $Pr_{1-x}Ca_xMnO_3$, IMT or colossal reduction of resistivity can be induced by certain external factors other than magnetic field, such as photon illumination, [13-16] electric current [17,18] and pressure.[19,20] These phenomena seem unpredictable in the DE framework. Third, the effects of disorder were also investigated theoretically. It is predicted that the quenched disorder suppresses the charge/lattice ordering and decreases the resistivity considerably, while the ferromagnetic ordering remains much less affected.[21-23] Recent experiments [24] evidenced this prediction: the A-site-disorder in $Pr_{0.6}R_{0.1}Sr_{0.3}MnO_3$ (R = Tb, Y, Ho, and Er) induced by different R-doping indeed causes significant resistivity change, while the measured magnetization as a function of $T$ changes little. Finally, AFM metallic state in some manganites at a divalent doping $x \sim 0.5$, e.g. $La_{0.46}Sr_{0.54}MnO_3$ [25] and $Nd_{0.45}Sr_{0.55}MnO_3$ [26] was experimentally observed, which is incompatible with the DE framework. More inconsistently, the possible anisotropy of transport properties as suggested in A-type AFM structure by Kawano *et al*, [27] has so far not yet been observed, although the anisotropy of spin correlation has been repeatedly confirmed experimentally, for example, in $Pr_{0.5}Sr_{0.5}MnO_3$ and $Nd_{0.45}Sr_{0.55}MnO_3$ by neutron diffraction technique, noting that the A-type AFM is ferromagnetism in planes and antiferromagnetism between those planes.

These crucial experimental findings and theoretical arguments encourage us to re-clarify the correlation between conductivity and magnetism. The transport behaviours of manganites are essentially determined by spin correlation and charge correlation. Whatever intrinsic interactions or external fields are, their effects on the transport should be viewed as direct driving forces. Due to the DE process, nonzero angle between spins of nearest neighbor (NN) Mn cations restrains $e_g$ electron movement. We denote by $\rho_s$ this spin-dependent resistivity. As the magnetism of the system transits within PM/FM/AFM, the change of $\rho_s$ is certainly within limited orders of magnitude. In addition, it has been frequently verified that the system resistivity can be very different upon different charge configuration. The extra resisitvity is expressed as $\rho_c$, which depends on variation of the charge configuration. In general, we can write the total resistivity as $\rho=\rho_s+\rho_c$. In fact, the charge correlation is much more important than spin correlation in contributing to the colossal resistivity change. And in many cases, the conductivity is independent from the spin configuration, although the DE mechanism correlates the metallic conductance with FM in a few special objects.

Figure 1 shows a sketch map of the present framework. The red curve, starting from the CO transition temperature ($T_{CO}$), represents the $\rho$-$T$ dependence (CO curve) for a pure CO state in which no contribution from spin degree of freedom is argued (to be confirmed below), while the blue $\rho$-$T$ curve (FM curve) which is relatively weak $T$-dependent is for a system of PM-FM transitions in absence of CO. A CO state has very poor conductance decreasing rapidly with decreasing $T$, [17] while an FM ordering makes the conductance metallic due to the spin-dependent DE mechanism. In fact, Aliaga *et al* suggested that there are two kinds of CMR (CMR1 and CMR2) effects.[21] CMR1 is induced by a very abrupt first-order transition at low $T$, while CMR2 is the more standard CMR effect that appears in the regime around $T_C$. In our present framework, CMR2 occurs around $T_C$, as indicated by arrow 1. Here magnetoresistance (MR) is not very large because the FM curve changes smoothly with $T$. And the PM-FM transitions near $T_C$ are continuous in response to applied magnetic field. A corresponding typical example is $La_{1-x}Sr_xMnO_3$ ($x\sim0.3$) which is a metal over the whole $T$ range.[2] Different from CMR2, CMR1 is a consequence of high-field induced CO-FM transitions at low $T$, indicated by arrow 3. It requires a large critical magnetic field, and produces a really colossal reduction of resistivity.

In addition, a charge disordered (CDO) state without FM ordering is introduced in our framework, as shown by CDO curve. Such a CDO state can be a spin glass or an AFM state of no long range CO, which has been observed in $La_{0.23}Ca_{0.77}MnO_3$.[28] The zero-field transport of CDO state is insulated, but the resistivity has a much milder $T$-dependence than pure CO states and consequently at low $T$ it is at least a few orders of magnitude lower than that of the CO states. The CDO state can be referred as a bad insulator here. The transport behaviour of $Pr_{0.5}Sr_{0.5}MnO_3$ (no clear sign of CO from neutron diffraction) below Neel point $T_N$ (AFM, $T_N \sim 150K$) [27] can be considered as an example of the CDO curve. This CDO state is an important concept to understand the X-induced resistivity change (we call it XR), here X may be any physical process such as photon illumination, pressure, electric current, disorder and so on. The X process may melt the CO state into the CDO state without changing the spin ordering much.[18] As shown in Figure 2, the periodic charge density in CO states, or periodic Coulomb repulsion, is absent in CDO states, in which most Mn cations have the averaged $e_g$ electron density. Then it is advantaged for carriers moving in a relatively leveled potential background and the conductivity is largely increased (several orders of magnitude) in CO-CDO transition, as indicated by arrow 4 in Fig. 1. The CDO state can be easily transferred to FM state, and the resistivity does not change much although the system changes from a bad insulator to a bad metal, from the point of view of $T$-dependence. In fact, this CDO-FM transition may be triggered by a spin-dependent DE-mediated process, and thus it can be classified as a type of CMR2 (arrow 2 in Fig. 1). Although CO states are stable against magnetic field or pressure below the threshold, they are sensitive to the site disorder because of the cooperative lattice effect. [8] Local illumination [13-16] may have the same effect as local disorder. And in some experiments, illumination may cause IMT more than XR, because DE can not be neglected in these manganites.[13,14]

As an extension of the CMR1 definition of Aliaga *et al*, in our framework the CMR1 effect as indicated by arrow 3 in Fig.1 can be understood as two sequences: a CO state is melt into a CDO state, followed by a CDO-FM transition. Thus, one has CMR1=XR+CMR2 with XR>>CMR2, although XR originates from an insulator-to-insulator transition rather than an IMT. The partition of CMR1 into XR and CMR2 is not trivial, which helps to understand the true mechanism of CMR and develop novel devices for potential application of manganites. Here CMR2 corresponds to the change of $\rho_s$, while XR is determined by the change of $\rho_c$. The XR, which is usually colossal and induced by disturbances other than magnetic field, can be utilized for device operation, since the large magnetic field required for CMR effect and imbalance between high $T_C$ and large MR make magnetic applications of manganites hard to break through.[1]

To further investigate the XR process, we perform a preliminary simulation on a toy model with CO/CDO states (see Appendix for details). According to the above argument, no spin degree of freedom is necessary for XR generation. Thus, our simulation does not consider the DE mechanism and magnetism, where the NN Coulomb repulsion is responsible for the resistance. When disorder is introduced into the system, the Coulomb repulsion-induced CO state will be suppressed, generating the XR effect. The simulated result of electronic transport is shown in Fig.3. Although this model is far simple from the real complex manganites, the result agrees quite well with corresponding experimental results qualitatively.

The above picture can also be applied to explain those observed but not well explained phenomena. The AFM metallic behaviour in some manganites actually corresponds to a transition of a weak CO state to a CDO state of AFM order with decreasing *T*, as indicated by arrow 5 in Fig. 1. Furthermore, A-Type AFM manganites are CO/CDO states and their transport behaviours are featured by the CO/CDO curves, whereas the spin degree of freedom does not contribute to the transport. Therefore, no transport anisotropy is expected although the spins are layered parallel arranged in these manganites.

Recent years the concept of PS as one origin of CMR effect becomes very popular. The PS pattern on micrometer scale [2-4] indicates micrometer correlation length of the intrinsic inhomogeneity. Taking lattice site-disorder into account, the random chemical potential models become inapplicable to describe percolation because the correlation length of disorder in those models is on the lattice unit scale, much smaller than the percolation scale, although cooperative lattice effects could enlarge correlation region.[8] Therefore, some new approaches were developed recently.[30,31] Nevertheless, the random chemical potential models have been quite successful in interpreting many experimental observations,[7,8,22,23] emphasizing the essential role of the site-disorder. In our present framework, site-disorder on lattice scale can suppress CO states,[22] and the micrometer PS is not necessary for generating the XR or CMR2 effects. Our framework is supported by recent experiment in which the CMR effect without PS was observed and ascribed to the disorder-induced spin glass state.[32] However, existence of PS really makes the transport behaviours more complex,[2-4] and its origin is still open for further study.

In summary, several major issues as revealed by extensive experimental investigations on the transport properties of manganites cannot be reasonably understood in the current theoretical framework, and they have been highlighted here. An extended framework for electronic transport of manganites has been proposed, in which the total resistivity has been partitioned into two parts: $\rho_s$ and $\rho_c$ in terms of two different mechanisms. The competition between the two mechanisms has been successfully applied to explain a series of different transport behaviours of manganites. The CMR classification inherited from Aliaga *et al* has been extended to include the spin-independent XR. We emphasize the important role of XR which helps to understand the true mechanism of CMR and may shed lights on a further exploration of novel applications of manganites. In addition, we argue that the micrometer phase separation as a phenomenon in manganites may not be a necessary origin for generating CMR effect.

We thank E. Dagotto, F. Yuan, H. Yu for critical comments and suggestion. S. Dong thanks S. Dai for assistance. This work was supported by the Natural Science Foundation of China (50332020, 10021001) and National Key Projects for Basic Research of China (2002CB613303, 2004CB619004).

*Appendix:*

The following Monte-Carlo simulation is performed to study the electronic transport behaviours of CO/CDO states without magnetic mechanism concerned. We consider "non-spin" electrons in a three dimensions $N \times N \times N$ cubic lattice. The Hamiltonian of the model is:

$$H = \sum_{<i,j>} e\vec{E} \cdot (\vec{i} - \vec{j}) c_i^+ c_j + V \sum_{<i,j>} n_i n_j + \mu_i n_i$$

Here $e$ is the charge unit; $E$ is the applied electric field; $i$, $j$ are the coordinates of the neighbor sites; $c_i^+$ ($c_i$) is the electron creation (annihilation) operator and $n_i = c_i^+ c_i$ is the number operator; $V$ is the Coulomb repulsion between $e_g$ electrons on the NN sites $i$ and $j$; $\mu_i$ is the chemical potential of $e_g$ electrons; and the disorder value $\delta$ is defined as the average difference of chemical potential $\mu_i$ between the two NN sites. Electrons can hop between NN sites but the double electrons occupation of one site is forbidden due to large Coulomb repulsion (in real manganites, the Jahn-Teller effect gives similar result), so $n_i = 0$ or 1.

Beginning the MC simulation, we distribute about $N^3/2$ electrons in the $N^3$ lattice sites in an ordered way, corresponding to the half filled manganites. Then a standard *Metropolis* algorithm is employed. In each step, a site $i$ is selected at random. If $n_i=0$, we break out and start the next step. In the $n_i=1$ case, we suppose that the electron in site $i$ can hop to the neighbor site $j$ if $n_j=0$, and we compare the energy in different states before and after this hopping. Then we calculate the probability of this movement $p_{ij}$:

$$p_{ij} = \exp(-\frac{H_i - H_j}{k_B T})$$

We execute the electron hopping to neighbor hole sites according to the probability. And the non-zero electric field causes the $x$-direction electrical current.

We have to mention that the true CO pattern in manganites (checkerboard arrangement in X-Y plane and charge-stacking along Z axis) can not be stable without anti-ferromagnetic super exchange between NN $t_{2g}$ spins. However, we exclude the spin degree of freedom in our toy model, so the structure of charge order here is the electron Wigner crystal, which is also a CO state. We use the toy model for the reasons that the colossal XR process is unaffected by such simplification, since the difference of resistivity between these CO patterns is not dominant. Besides, this model can obtain the pure CO/CDO without spin's influence which makes the result too complex to analyze, as in real system case.

*Figure caption:*

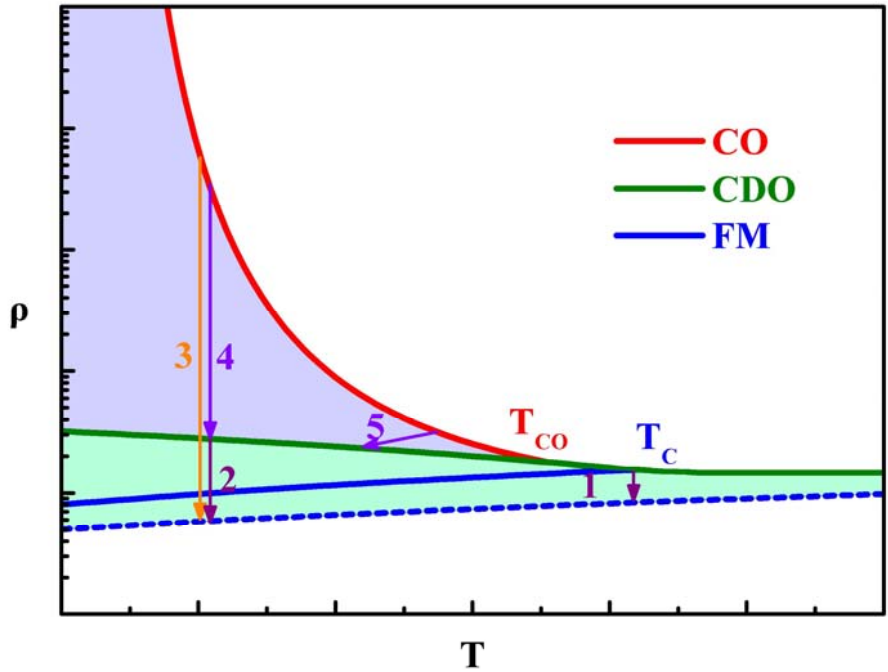

FIG. 1. The *T*-dependent transport behaviors of CO/CDO/FM states are plotted as red/green/ blue curves, respectively. The blue dashed curve represents the transport behavior of the FM state under a magnetic field. The vertical axis is resistivity in logarithm. The purple and green regions are dominated by variations of charge degree of freedom and spin degree of freedom, respectively.

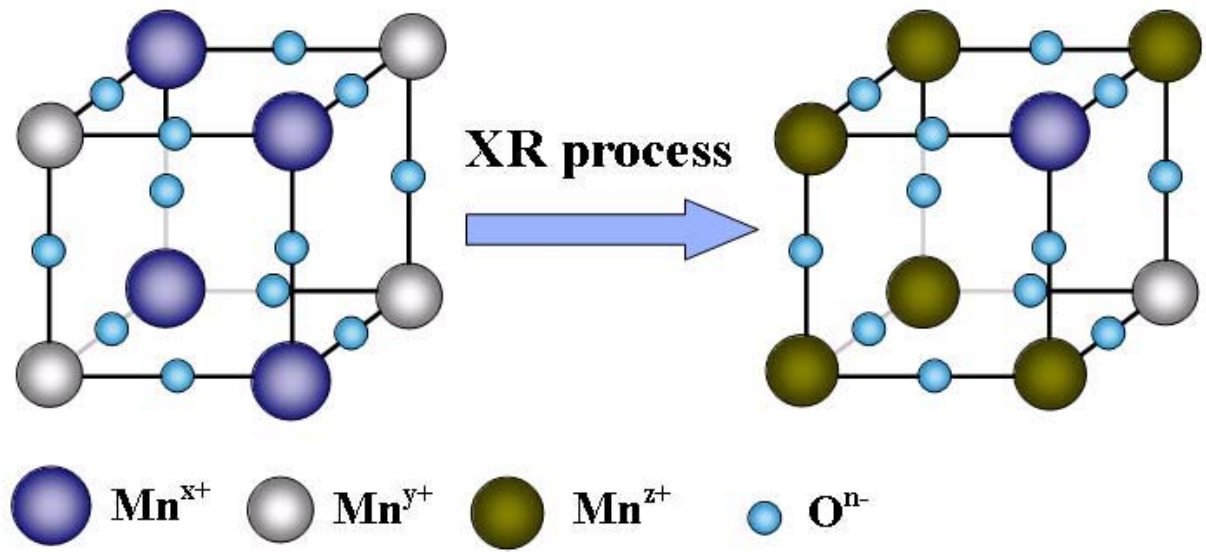

FIG. 2. Charge configuration of CO/CDO states. Here the arrangement of $e_g$ electron in the CO state is a charge stacking configuration. Here $Mn^{x+}$, $Mn^{y+}$, $Mn^{z+}$ and $O^{n-}$ are $Mn^{3+}$, $Mn^{4+}$, $Mn^{3.5+}$ and $O^{2-}$ respectively in common knowledge. Recently, it is argued that the valences of Mn and O ions are more complex in real CO states. [29] However, it does not affect our framework because the period of charge density still exist although $y-x \neq 1$.

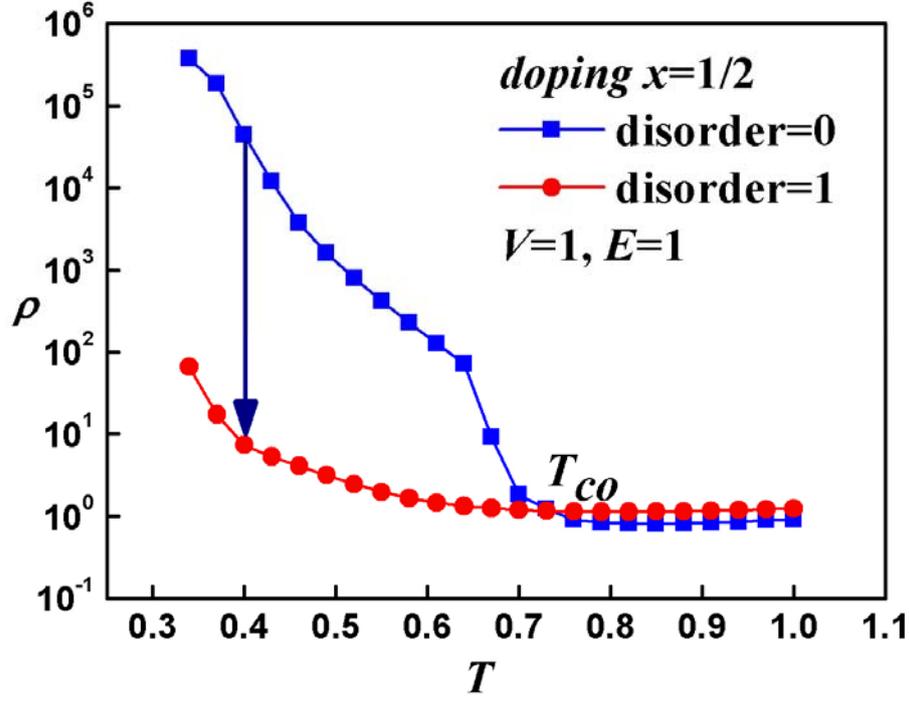

FIG. 3. Simulated $T$-dependent resistivity. All parameters and results here are relative values. When all $\mu_i$ equal to $\mu$ ($\delta=0$), the resistivity increases rapidly while $T$ decreases below $T_{CO}$. When each $\mu_i$ is set as a random value, e. g. $\delta=1$, with cooperative lattice effects considered, [7] the resistivity reduction is up to several orders of magnitude in low $T$: a XR process indicated by the arrow. At high $T$, the resistivity increases due to the disorder, as compared with the low $T$ case, which is accordance with the basic concept of Anderson transition.